\documentclass{Interspeech2024}

\usepackage{amsmath, amssymb, amsfonts}
\usepackage{cite}

\interspeechcameraready

\title{VAE-based Phoneme Alignment Using Gradient Annealing and \\
SSL Acoustic Features}

\name[affiliation={1}]{Tomoki}{Koriyama}

\address{
  $^1$CyberAgent, Japan
\email{koriyama\_tomoki@cyberagent.co.jp}
}


\begin{document}

\maketitle

\begin{abstract}
This paper presents an accurate phoneme alignment model that aims for speech analysis and video content creation. We propose a variational autoencoder (VAE)-based alignment model in which a probable path is searched using encoded acoustic and linguistic embeddings in an unsupervised manner. Our proposed model is based on one TTS alignment (OTA) and extended to obtain phoneme boundaries. Specifically, we incorporate a VAE architecture to maintain consistency between the embedding and input, apply gradient annealing to avoid local optimum during training, and introduce a self-supervised learning (SSL)-based acoustic-feature input and state-level linguistic unit to utilize rich and detailed information. Experimental results show that the proposed model generated phoneme boundaries closer to annotated ones compared with the conventional OTA model, the CTC-based segmentation model, and the widely-used tool MFA.
\end{abstract}

\section{Introduction}

Alignment in speech processing refers to the task of determining the times at which phonemes, characters, or words are uttered, based on a pair of a speech waveform and a phoneme or character sequence.
Accurate alignment can be used to analyze phonetic and prosodic features of speech in detail, as well as to
create video content such as subtitles, lip-sync, and speech editing.
Furthermore, the alignment between the speech signal and the phoneme or text is crucial for modeling speech recognition and synthesis, because these tasks involve sequence-to-sequence transformations with different lengths.

The most well-known conventional method of phoneme alignment is the use of the Gaussian mixture model (GMM)-based hidden Markov model (HMMs) \cite{rabiner1975applications},
which is also used the recent tool, Montreal Forced Aligner (MFA) \cite{mcauliffe2017montreal}.
To overcome the model expressiveness of GMM,
many neural network (NN)-based methods have been proposed \cite{kurzinger2020ctc,bain23whisperx,li2022neufa,chung21_interspeech,kelley18_interspeech,zhu2022phonetoaudio,kouzelis23_interspeech,raissi2023hmm,badlani2022one,shih2021rad,kim2020glowtts,kim2021conditional},
from the perspective of automatic speech recognition (ASR) and text-to-speech (TTS).
CTC segmentation \cite{kurzinger2020ctc} utilizes an ASR network trained with connectionist temporal classification (CTC) loss that includes a blank character.
Since CTC segmentation only requires ASR models,
it is easy to adapt various languages as WhisperX does \cite{bain23whisperx}.
However, the blank symbol conceals the beginning and end of each character, resulting in inaccurate boundary prediction.
Another approach uses supervised training to predict the boundaries of phonemes \cite{li2022neufa}.
The issue with using a supervised method is that it is applicable to various languages or speaking styles because phoneme boundary annotation for training is very limited.

TTS research also developed alignment model to capture the relationship between text and waveform.
Famous frameworks use a forward-sum algorithm \cite{shih2021rad,badlani2022one} or its Viterbi search variant, known as monotonic alignment search (MAS) \cite{kim2020glowtts}, which searchs for a monotonic path with dynamic programming.
For example, Glow-TTS \cite{kim2020glowtts} and VITS \cite{kim2021conditional} perform MAS using acoustic embeddings extracted by normalizing flow architecture,
while RAD-TTS \cite{shih2021rad} and one tts alignment (OTA) \cite{badlani2022one}
utilize forward-sum loss with NN-based linguistic and acoustic encoders.
The use of forward-sum loss enables simultaneous training of alignment and TTS.
However, the phoneme boundaries obtained from alignment modules are not always accurate,
even though the synthetic speech is natural sounding due to the high expressiveness of NN models.
Moreover, the accuracy of alignment has not been precisely evaluated in the previous studies \cite{kim2021conditional,badlani2022one}.

In this study, we explore phoneme alignment that can predict accurate boundaries in an unsupervised manner.
Specifically, we propose a novel alignment model by extending the alignment module of OTA \cite{badlani2022one}.
We incorporate the variational autoencoder (VAE) \cite{kingma2013auto} architecture to retain time-dependent acoustic and linguistic information of encoders.
We also introduce acoustic features obtained from self-supervised learning (SSL) models \cite{baevski2020wav2vec,hsu2021hubert} to utilize rich acoustic information,
similar to recent alignment studies \cite{bain23whisperx,zhu2022phonetoaudio}.
Furthermore, we investigate the use of gradient annealing to avoid local optima during the path search, inspired by the gradient-based GMM training technique \cite{gepperth2021gradient}.

Experimental evaluations are conducted using a corpus that include manually-annotated phoneme boundaries.
We calculate the errors of boundaries predicted by various modeling settings
and demonstrate that the proposed method can yield more accurate phoneme boundaries compared with the conventional methods.
Moreover, we perform an ablation study to examine the effectiveness of each component of the proposed model.
Some examples are available at \footnote{\url{https://github.com/hyama5/vae_align}}.

\section{One TTS Alignment}

A popular method for text or phoneme alignment involves
searching for an appropriate path between an acoustic
embedding sequence $(\mathbf{y}_1, \dots, \mathbf{y}_T)$
and a linguistic feature sequence $(\mathbf{x}_1, \dots, \mathbf{x}_K)$.
The parallel version of OTA \cite{badlani2022one}, which we refer to as OTA in this paper and is derived from RAD-TTS \cite{shih2021rad}, uses convolutional networks to obtain the embeddings
and trains the network by minimizing the following forward-sum loss function:
\begin{align}
L_\mathrm{align} = -\log \sum_{\mathbf{s} \in S} \prod_{t=1}^T b(t, s_t).
\end{align}
Here, $\mathbf{s} = (s_1, \dots, s_T) ~(s_t \in \{1, \dots, K\})$ is the state sequence and
$S$ is a set of monotonic left-to-right no-skip paths.
The function $b(t, s_t) > 0$ represents the likelihood that the acoustic feature at time $t$ has the $s_t$-th linguistic feature,
which is defined by the product of a matching function $f(\mathbf{y}_t, \mathbf{x}_k)$ and a position prior $p_\mathrm{pos}(t, k; \omega)$ as follows:
\begin{align}
b(t, k) = f(\mathbf{y}_t, \mathbf{x}_k) p_\mathrm{pos}(t, k; \omega).
\label{eq:likelihood}
\end{align}
The matching function $f(\cdot)$ is a softmax function with the logits of negative squared distance
given by 
\begin{align}
f(\mathbf{y}_t, \mathbf{x}_k) &= 
\frac{\exp(-\lVert \mathbf{y}_t - \mathbf{x}_k \rVert^2)}
{\sum_{i=1}^K \exp(-\lVert \mathbf{y}_t - \mathbf{x}_i \rVert^2)}.
\end{align}
The position prior $p_\mathrm{pos}(\cdot)$ is a bias function defined independently from acoustic and linguistic features
to represent the rough diagonal tendency of paths.
Specifically, the position prior is implemented using a beta-binomial distribution
$p_\mathrm{pos}(t, k; \omega) = \mathrm{BetaBin}(k, \omega t, \omega (T - t + 1))$.
Although a large value of $\omega$ helps fast convergence during training,
it can also impose a strong restriction because not all speech utterances have diagonal paths.

The training of encoders and decoders is performed by gradient-based optimization
using the forward-sum loss $L_\mathrm{align}$.
The gradient for $\log b(\cdot)$ is given by
\begin{align}
\frac{\partial L_\mathrm{align}}
{\partial \log b(t, k)}
= 
-\frac{\sum_{\mathbf{s} \in S(t,k)} \prod_{\tau=1}^T b(\tau, s_\tau)}
      {\sum_{\mathbf{s} \in S} \prod_{\tau=1}^T b(\tau, s_\tau)}
\triangleq -\gamma(t,k)
\label{eq:gradient}
\end{align}
where $S(t,k)$ is the set of paths through state $k$ at time $t$.
The term $\gamma(t,k)$ represents the probability of passing through $(t, k)$,
which is known as state occupancy in the training of HMM.
It is possible to compute the $\gamma(t,k)$ using
the dynamic programming of the forward-backward algorithm in a similar manner to HMM.
When predicting phoneme boundaries from the trained model,
we perform a Viterbi search, also known as monotonic align search,
to find the most likely path
and identify the state for each time.

\section{Proposed VAE-based Alignment}

We propose a novel alignment model by extending OTA.
The major modifications from OTA include a VAE-based architecture, 
gradient annealing, SSL acoustic features, and state-level linguistic units.

\begin{figure}[t]
  \centering
  \includegraphics[width=\linewidth]{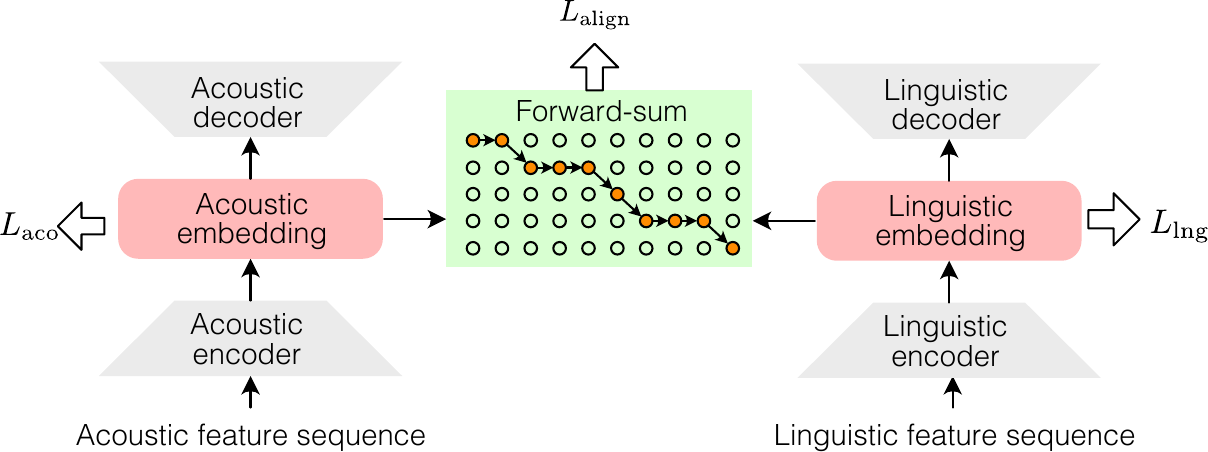}
  \caption{Architecture of the proposed alignment model.}
  \label{fig:model}
\end{figure}

For accurate alignment,
the acoustic embedding of the $t$-th frame should retain the characteristics of the acoustic feature of the $t$-th frame, and the same applies to linguistic embeddings.
However, we have observed that when we use the OTA model, the acoustic and linguistic embeddings do not always
maintain the original acoustic and linguistic information
and often overfit the alignment loss.
To address this issue, we incorporate a VAE architecture \cite{kingma2013auto} as shown in Fig.~\ref{fig:model}.
We expect that this architecture will preserve the input acoustic and linguistic information
in the embedding layers.
The loss function is defined by
\begin{align}
L = L_\mathrm{align} + w_\mathrm{aco} L_\mathrm{aco} +  w_\mathrm{lng} L_\mathrm{lng}
\end{align}
where $w_\mathrm{aco}$ and $w_\mathrm{lng}$ are weights, and $L_\mathrm{aco}$ and $L_\mathrm{lng}$ are VAE losses, respectively.

Moreover, gradient-based training often falls into a local optimum.
That is, the path becomes fixed in the early stages of training,
making it difficult to search for the probable path during training. 
This issue aises from the gradient given in Eq.~(\ref{eq:gradient}).
If the likelihood $b(t, k)$ is very small,
the gradient becomes close to zero,
and thus the parameters for the likelihood $b(t, k)$ are slightly updated.
This phenomenon is observed in similar models such as HMM or GMM
that have hidden states when we perform gradient-based optimization \cite{gepperth2021gradient}.
Therefore, we propose annealing for the gradients in the alignment, based on the annealing in GMM training \cite{gepperth2021gradient}.
Specifically, we employ convolution with a Gaussian annealing filter $g(\cdot)$ given by
\begin{align}
\frac{\partial L_\mathrm{align}}
{\partial \log b(t, k)}
&\approx -\gamma'(t,k) = -\gamma(t,k) * g(k) \\
g(k) &= \exp\left(-\frac{k^2}{2\sigma^2}\right)
\label{eq:approx_gradient}
\end{align}
where $\sigma$ is the annealing coefficient, and $*$ denotes convolution.
The coefficient $\sigma$ is updated by multiplying a rate $r < 1.0$ during training.
The annealed gradient approximates the original one in Eq.~(\ref{eq:gradient}) when $\sigma$ becomes sufficiently small.
This annealing shares the possibility with adjacent states,
which helps to avoid the local optimum problem.

The use of SSL models has had a significant impact on the field of speech processing.
For example in alignment, the CTC segmentation demo \cite{wav2vec_ctc} and WhisperX \cite{bain23whisperx} utilize SSL models to perform alignment.
We apply SSL features extracted from a pretrained SSL model to the input of acoustic features.
Since SSL models are trained by predicting the masked frame-level components,
the SSL features can provide position-dependent rich information that is useful for accurate alignment.

We also examine the alignment level of linguistic units.
In OTA \cite{badlani2022one} and other models \cite{shih2021rad,kim2021conditional,kurzinger2020ctc}, phonemes or characters are used as the unit of alignment.
This assumes that each phoneme or character has one stable feature in acoustic embeddings.
However, some phonemes change drastically within themselves.
For example, a plosive sound has two different segments: a silent part and a bursting part, which is often referred to as subphone.
Therefore, we extend the linguistic unit to the multiple-state level
by dividing the linguistic input of one phoneme into multiple states per phoneme.
This approach allows us to capture the changes within the phonemes,
similar to the GMM-HMM-based methods.

\section{Experiments}

\subsection{Experimental conditions}

\subsubsection{Dataset}

We used the Corpus of Spontaneous Japanese (CSJ) \cite{Maekawa_2003} for experimental evaluations.
The core data of CSJ, which includes detailed phoneme annotation,
was divided into training, development, and evaluation subsets
based on the ASR recipe of ESPnet \cite{watanabe2018espnet}.
The training, development, and evaluation subsets comprised
24,863 utterances (188 talks, $~$1,580 k phonemes, $~$33.2 h),
900 utterances (8 talks, $~$56 k phonemes, $~$1.2 h),
896 utterances (8 talks, $~$62 k phonemes, $~$1.3 h),
respectively.
We used 52 entities as a phoneme set
and regarded the annotated phoneme boundaries as the true ones for evaluation.
It should be noted that these annotated boundaries were not used during training.

\subsubsection{Acoustic features}

We used mel-frequency cepstral coefficient (\textbf{MFCC}) and melspectrogram (\textbf{Melspec})
as conventional acoustic features.
MFCC was a 39-dimensional feature consisting of 13-dimensional coefficients and their $\Delta$ and $\Delta^2$ features.
The dimension of melspectrogram was set to 80.
The acoustic features were extracted with a 10 ms frame-shift.
For SSL features, we used four types of SSL-based models trained with different objectives: \textbf{HuBERT(en)}, \textbf{HuBERT(ja)},
\textbf{XLS-R}, and \textbf{w2v2-ASR}.
HuBERT(en) was a HuBERT-base model \cite{hsu2021hubert} trained with English Librispeech \cite{panayotov2015librispeech} database,
whereas HuBERT(ja) was a HuBERT-base model trained with the Japanese ReazonSpeech \cite{reazonspeech} database \cite{rinna_hubert}, which is the same language as the target one.
XLS-R was a wav2vec2.0 large model \cite{babu2021xls} trained with multiple languages.
The model w2v2-ASR was fine-tuned from XLS-R using Japanese corpora.
We also used Whisper encoder \cite{radford2023robust} (\textbf{Whisper}), which has recently gained popularity as a base model for feature extraction. 
Since SSL and Whisper features were defined with a 20 ms shift,
we upsampled them to a 10 ms shift to compare with the conventional features.

\subsubsection{Base model configurations}

We define a base proposed model used for the following evaluations.
The encoders and decoders for acoustic and linguistic features were composed of 6-layer convolution networks, respectively.
The kernel size of convolution layers was 3 and 256 hidden channels were employed.
The dimension of embeddings for the matching function was 64.
We adopted 3 states per phoneme as the modeling level of linguistic units.
The loss weights were set as $(w_{\mathrm{aco}}, w_{\mathrm{lng}}) = (0.1, 0.1)$.
The coefficient of position prior $\omega$ was set to 0.01.
We set the initial annealing coefficient $\sigma$ and update rate $r$ to 30.0 and 0.9, respectively,
and update $\sigma$ every 1000 training steps.
The training was performed up to 90k steps, 
and using Adam optimization \cite{kingma2014adam} with a learning rate of $1.0 \times 10^{-5}$.
The minibatch size was 4.

We implemented the computation of dynamic programming for the forward-sum loss using
Triton \cite{tillet2019triton} and other computations using PyTorch.
Training time with one NVIDIA A100 GPU took approximately 150 min.

\subsubsection{Conventional methods}

We used MFA, CTC segmentation, and OTA for comparison.
The MFA was trained using the same dataset as the other models
utilizing the default configuration of MFA.
To create the CTC-segmentation model,
we trained a phoneme recognition model composed of a 6-layer Conformer \cite{gulati20conformer} with CTC loss.
As the input of the model, we used the output of HuBERT(ja) L12.
The configuration of the OTA model was based on the original OTA paper \cite{badlani2022one}.
Specifically, we used two and one convolution layers for acoustic and linguistic encoders, respectively.
The acoustic features were melspectrogram, and
the modeling level of linguistic features was a phoneme.
In this study, only alignment loss was used instead of TTS loss for the OTA model.
We used the same configurations as the proposed base model for other experimental settings.

\subsubsection{Evaluation metrics}

We evaluated the performance of the alignment models
by calculating the errors between the predicted and annotated phoneme boundaries.
There is no consensus on the accuracy of phoneme boundaries, which depends on the granularity of words and phonemes to be analyzed.
Hence, we used the four evaluation metrics.
\textbf{MAE} and \textbf{Median} were the mean and median of absolute errors. MAE was more sensitive to outliers than Median.
The other metrics were tolerance error rates
defined as the ratio of errors whose absolute values exceeded a specific threshold.
We used 20 ms and 50 ms as the threshold.
All values in the evaluation, except for MFA,
were the average of 5 trials with different random seeds.

\begin{table}[t]
  \caption{Mean and median of boundary errors [ms] and tolerance error rates [\%]
  on different acoustic features.}
  \footnotesize
  \label{tab:acoustic_features}
  \centering
  \begin{tabular}{ r | rr rr }
    \toprule
    Features & MAE & Median & 20 ms tol & 50 ms tol\\
    \midrule
    MFCC & 15.29 & 10.24 & 21.0 & 3.57 \\
    Melspec & 15.46 & 10.59 & 21.8 & 3.20 \\
    \midrule
    HuBERT(en) L0 & 14.13 & 8.49 & 18.3 & 3.45 \\
    HuBERT(en) L4 & 13.33 & 8.67 & 18.0 & 2.45 \\
    HuBERT(en) L8 & 13.53 & 9.02 & 18.0 & 2.40 \\
    HuBERT(en) L12 & 13.53 & 9.03 & 18.4 & 2.40 \\
    \midrule
    HuBERT(ja) L0 & \textbf{12.91} & \textbf{8.25} & \textbf{16.1} & 2.59 \\
    HuBERT(ja) L4 & 13.40 & 8.69 & 17.8 & 2.56 \\
    HuBERT(ja) L8 & 14.05 & 10.46 & 19.8 & \textbf{2.13} \\
    HuBERT(ja) L12 & 13.40 & 9.33 & 17.7 & 2.21 \\
    \midrule
    XLS-R L0 & 13.66 & 8.50 & 17.4 & 3.16 \\
    XLS-R L8 & 13.68 & 8.50 & 17.4 & 3.16 \\
    XLS-R L16 & 13.66 & 8.50 & 17.3 & 3.17 \\
    XLS-R L24 & 13.79 & 8.92 & 17.9 & 2.75 \\
    \midrule
    w2v2-ASR L0 & 13.62 & 8.56 & 17.3 & 3.08 \\
    w2v2 ASR L8 & 14.46 & 9.93 & 21.0 & 2.83 \\
    w2v2-ASR L16 & 17.92 & 12.56 & 30.4 & 4.08 \\
    w2v2-ASR L24 & 17.33 & 12.05 & 28.6 & 4.58 \\
    \midrule
    Whisper & 22.89 & 19.54 & 48.6 & 4.93 \\
    \bottomrule
  \end{tabular}
  
\end{table}

\subsection{Results on acoustic features}

We first evaluated the effect of acoustic features on phoneme alignment.
Table~\ref{tab:acoustic_features} shows the results for each feature. 
In the table, L0 to L24 represent the layer indices of intermediate transformer layers.
When we compare the conventional features of MFCC and Melspec with
HuBERT and XLS-R, HuBERT and XLS-R features had smaller MAEs.
Since the differences among HuBERT(en), HuBERT(ja), and XLS-R
were not significant, the dependency of language and SSL modeling method
may not be crucial.
It is observed that w2v2-ASR had large errors when we used layers L16 and L24.
A possible reason is that the fine-tuned model for ASR lost the position information,
leading to worse alignment accuracy.
This is supported by the large errors of Whisper encoders trained for speech recognition objectives.
The following experiments were performed using the best features, \textbf{HuBERT(ja) L0}.

\begin{table}[t]
  \caption{Mean and median of boundary errors [ms] and tolerance error rates [\%]
  on different alignment methods.}
  \footnotesize
  \label{tab:methods}
  \centering
  \begin{tabular}{ r | rr rr }
    \toprule
    Method & MAE & Median & 20 ms tol & 50 ms tol\\
    \midrule
    MFA & 16.46 & 10.45 & 23.1 & 4.31 \\
    CTC & 40.81 & 34.63 & 70.9 & 30.87 \\
    OTA & 23.10 & 14.55 & 36.1 & 7.28 \\
    Proposed & 12.91 & 8.25 & 16.1 & 2.59 \\
    \bottomrule
  \end{tabular}
\end{table}

\subsection{Results on methods}

We compared the proposed method with the conventional ones
shown in Table~\ref{tab:methods}.
We see that CTC segmentation had larger errors than the other methods.
This could be because the existence of blank symbols and the CTC loss did not assure the accurate position information.
Although we tried multiple types of input features and encoder architectures,
the same tendency was observed.
We also confirmed that the proposed method, an extension of OTA model, reduced errors,
indicating that the extension worked as expected.
Although the GMM-HMM-based MFA yielded small errors,
the proposed method outperformed MFA.

\begin{table}[t]
  \caption{Mean and median of boundary errors [ms] and tolerance error rates [\%]
  on ablation of model components.}
  \footnotesize
  \label{tab:ablation}
  \centering
  \begin{tabular}{ l | rr rr }
    \toprule
    Method & MAE & Median & 20 ms tol & 50 ms tol\\
    \midrule
    Proposed & 12.91 & 8.25 & 16.1 & 2.59 \\
    w/o VAE & 14.96 & 10.41 & 23.1 & 2.90 \\
    w/o annealing & 13.26 & 10.45 & 16.8 & 2.89 \\
    Phoneme-level & 16.09 & 10.29 & 22.9 & 4.22 \\
    w/o position prior & 12.96 & 8.24 & 16.1 & 2.61 \\
     \bottomrule
  \end{tabular}
\end{table}

\subsection{Ablation study on proposed components}

The proposed method extends OTA models by adding multiple features.
Hence, we performed an ablation study to evaluate the importance of each component.
\begin{itemize}
\item \textbf{W/o VAE} only used $L_{\mathrm{align}}$ for training instead of $L$.
\item \textbf{W/o annealing} used the gradient of Eq.~(\ref{eq:gradient}) directly for training.
\item \textbf{Phoneme-level} employed phonemes as the linguistic unit instead of three states per phoneme.
\item \textbf{W/o position prior} did not use position prior of Eq.~(\ref{eq:likelihood}). This ablation was an evaluation of the position prior used in the OTA model.
\end{itemize}

The results are shown in Table~\ref{tab:ablation}.
It is seen that the proposed VAE, gradient annealing, and state-level modeling were all
important to reduce errors.
The position prior did not significantly affect the errors.
We observed that while the position prior was an important factor in convergence, since annealing was also effective for convergence, the position prior was not always necessary.

\begin{table}[t]
  \caption{Mean and median of boundary errors [ms] and tolerance error rates [\%]
  on ablation of encoders/decoders.}
  \footnotesize
  \label{tab:enc_dec}
  \centering
  \begin{tabular}{ r | rr rr }
    \toprule
    Method & MAE & Median & 20 ms tol & 50 ms tol\\
    \midrule
    Proposed base & 12.91 & 8.25 & 16.1 & 2.59 \\
    Conformer aco. enc.  & 12.93 & 8.31 & 16.0 & 2.65 \\
    Conformer aco. dec.  & 12.84 & 7.86 & 15.5 & 2.75 \\
    Conformer ling. enc. & \textbf{12.51} & \textbf{8.21} & \textbf{15.4} & \textbf{2.32} \\
    Conformer ling. dec. & 13.18 & 8.51 & 16.5 & 2.69 \\
    \midrule
    Flow aco. enc & 14.36 & 9.52 & 20.1 & 3.26 \\
    \bottomrule
  \end{tabular}
\end{table}

\subsection{Ablation study on encoder and decoders}

The encoders and decoders of base models were simply composed of 6 convolution layers.
However, we can consider using more complex and highly expressive components.
In this ablation study, we replaced the convolution layers of each acoustic/linguistic encoder or decoder with a 6-layer Conformer block \cite{gulati20conformer}, which included 4-head self-attention and depth-wise convolution modules.
Another ablation involved using normalizing flow for acoustic encoder instead of VAE as in VITS \cite{kim2021conditional}.
We used an affine coupling layer in the same way as VITS,
and trained the model of 6 flow steps and a 1-layer convolution for each flow step.

The results are shown in Table~\ref{tab:enc_dec}.
It is seen that the use of Conformer for the linguistic encoder reduced errors compared with the proposed base model.
A possible reason is that long-range linguistic contexts obtained by the Conformer could effectively
extract linguistic embeddings for alignment.
The errors using the flow-based acoustic encoder were larger than the proposed base model.
This could be because of the structural limitation that flows are required to be invertible.

\section{Conclusions}

In this study, we proposed an unsupervised phoneme alignment model based on the OTA model.
Specifically, we incorporated a VAE architecture to maintain consistency between inputs and embeddings,
applied gradient annealing to avoid local optima,
and incorporated SSL acoustic feature input and state-level linguistic unit
to utilize rich and detailed information.
In experimental evaluations, we measured alignment accuracy using annotated phoneme boundaries
and demonstrated that the proposed method can generate more accurate alignment than conventional methods, including GMM-HMM-based ones.
In future work, we should evaluate the effectiveness of the proposed model using multiple languages including low-resource ones and diverse styles such as
expressive speech and singing voices.
We will also investigate the effectiveness of the accurate alignment for practical applications such as TTS and video content creation.

\bibliographystyle{IEEEtran}
\bibliography{mybib}

\end{document}